\documentclass{article}

\usepackage{PRIMEarxiv}

\usepackage[utf8]{inputenc} 
\usepackage[T1]{fontenc}    
\usepackage{hyperref}       
\usepackage{url}            
\usepackage{booktabs}       
\usepackage{amsfonts}       
\usepackage{nicefrac}       
\usepackage{microtype}      
\usepackage{lipsum}
\usepackage{fancyhdr}       
\usepackage{graphicx}       
\graphicspath{{media/}}     

\usepackage[figuresright]{rotating}
\usepackage{tablefootnote}
\usepackage{subfig}
\usepackage{graphicx}
\usepackage{multirow}
\usepackage{soul}
\usepackage{enumitem}
\usepackage{adjustbox}
\usepackage{booktabs}

\usepackage{tikz}
\usetikzlibrary{positioning}
\usepackage{pgfplots}
\usetikzlibrary{matrix}

\usepgfplotslibrary{groupplots}
\usepackage{appendix} 
\usepackage{amsmath}

\title{A Simple Yet Effective Approach for Diversified Session-Based Recommendation
\thanks{\textit{\underline{Citation}}: 
\textbf{}} 
}

\author{
  Qing Yin, Hui Fang \\
  Shanghai University of Finance and Economics \\
  Shanghai, China\\
  \texttt{qyin.es@gmail.com, fang.hui@mail.shufe.edu.cn} \\
   \And
  Zhu Sun \\
  Institute of High Performance Computing; Centre for Frontier AI Research, A*STAR \\
  Singapore\\
  \texttt{sunzhuntu@gmail.com} \\
  \AND
  Yew-Soon Ong \\
  A*STAR Centre for Frontier AI Research; Nanyang Technological University \\
  Singapore \\
  \texttt{asysong@ntu.edu.sg} \\
}

\begin{document}
\maketitle

\begin{abstract}
Session-based recommender systems (SBRSs) have become extremely popular in view of the core capability of 
capturing short-term and dynamic user preferences. However, most SBRSs primarily maximize recommendation accuracy but ignore users' minor preferences,
thus leading to filter bubbles in the long run. Only a handful of works, being devoted to improving diversity, depend on unique model designs and calibrated loss functions, which cannot be easily adapted to existing accuracy-oriented SBRSs. 
It is thus worthwhile to come up with a \textit{simple yet effective} design that can be used as a plugin to facilitate existing SBRSs on generating a more diversified list in the meantime preserving the recommendation accuracy.
In this case, we propose an end-to-end framework applied for every existing representative (accuracy-oriented) SBRS, called diversified category-aware attentive SBRS (DCA-SBRS), to boost the performance on recommendation diversity. 
It consists of two novel designs: a model-agnostic diversity-oriented loss function, and a non-invasive category-aware attention mechanism. 
Extensive experiments on three datasets showcase that our framework helps existing SBRSs achieve extraordinary performance in terms of recommendation diversity (e.g., an average of 74.1\% increase on ILD$@$10) and comprehensive performance 
{(e.g., an average of 52.3\% lift on F-score$@$10)}, without significantly deteriorating recommendation accuracy compared to state-of-the-art accuracy-oriented SBRSs.
The source code can be obtained via \url{github.com/qyin863/DCA-SBRS}.
\end{abstract}

\maketitle

\section{Introduction}
Session-based recommender systems (SBRSs) have gained significant attention because they provide more timely and accurate recommendations by incorporating short-term and dynamic user preferences~\cite{fang2020deep,wang2021survey}.
To enhance recommendation accuracy, existing SBRSs utilize sophisticated models like deep neural networks that capture short-term preferences from the most recent session. For instance, GRU4Rec~\cite{hidasi2015session} employs gated recurrent units (GRU) to learn a session's sequential behaviors. Furthermore, the attention mechanism is imported to capture main-purpose (intent) preferences such as NARM~\cite{li2017neural} and STAMP~\cite{LiuZMZ18}. Moreover, 
graph neural networks (GNNs) are utilized to learn more complex item relationships (e.g., SR-GNN~\cite{WuT0WXT19}, GC-SAN~\cite{XuZLSXZFZ19}, and GCE-GNN~\cite{wang2020global}). 
For the above state-of-the-art (SOTA) SBRSs, \emph{attention mechanisms} are used together with RNNs or GNNs to improve recommendation performance~\cite{wang2021survey}. 
 
{However, the aforementioned  
SOTA (\emph{accuracy-oriented}) SBRSs would gradually overemphasize dominant interests and weaken minor ones \cite{steck2018calibrated}, thus leading to a filter bubble~\cite{nguyen2014exploring,khenissi2020theoretical} over time.}
{As such,}
\emph{diversified} recommender systems (RSs) are raised to recommend more diverse lists (e.g., with items covering many categories). The diversified works in traditional recommendation fall into three major categories: post-processing heuristic methods~\cite{carbonell1998use,steck2018calibrated}, determinantal point process (DPP) methods~\cite{chen_fast_2018,wu2019pd,gan2020enhancing} and end-to-end learning methods~\cite{zheng2021dgcn,liang2021enhancing}.
However, to the best of our knowledge, there are only three representative diversified SBRSs such as MCPRN \cite{Wang0WSOC19}, ComiRec \cite{Cen2020ControllableMF} and IDSR \cite{chen2020improving}. Both MCPRN and ComiRec design multiple channels rather than one major channel to learn multiple purposes in a session, where recommendations strive to satisfy these purposes instead of only capturing the main purpose as representative accuracy-oriented works (e.g., NARM). Following the above multiple-purpose assumption, IDSR also jointly incorporates both item relevance and diversity into the prediction score and loss function. 

\begin{figure}[t]
    \centering
    \includegraphics[width=.4\textwidth]{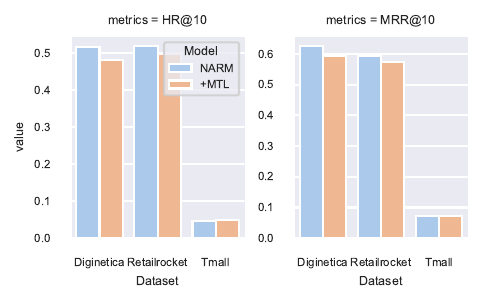}
    \vspace{-0.15in}
    \caption{
    NARM vs NARM+MTL. Note: +MTL denotes the variant of NARM via leveraging item categories as input and adopting the common multi-task learning framework.
    }
    \label{fig:mtl_compar}
    \vspace{-0.1in}
\end{figure}

To conclude, existing studies on diversified SBRSs mainly suffer from two challenges: (1) as we can tell from previous studies, model variants like multiple channels and unique diversity-oriented loss (objective) fitted for special diversity modules are carefully calibrated by diversified SBRSs. However, such diversified designs cannot be easily adopted by existing representative accuracy-oriented SOTA SBRSs.
Thus, the first research challenge lies in how to come up with simple yet effective designs (like loss function) that can facilitate the diversity performance of SOTA accuracy-oriented SBRSs?
and (2) previous diversified works mostly fail to obtain a comparable performance on accuracy to those representative accuracy-oriented SBRSs, since in most cases improved diversity is reached at the cost of sacrificing a certain level of accuracy. To mitigate the adversarial effect, side information like category of items is generally imported to help better learn user preferences~\cite{zhao2018categorical,sun2019research,liu2021noninvasive}. However, for representative accuracy-oriented SBRSs, we surprisingly find that simply concatenating item ID and its category information as the input and adopting the
common multi-task learning framework, as in SBRS+MTL~\cite{zhao2018categorical}, cannot considerably improve recommendation performance and may even result in worse performance in terms of accuracy metrics (see Figure~\ref{fig:mtl_compar}). 
In this case, our second challenge is to seek for a solution that can help maintain recommendation accuracy for diversified SBRSs by better exploiting category information.

Towards the aforementioned two issues, we propose a simple yet effective end-to-end Diversified Category-aware Attentive framework that can be easily instantiated with existing representative accuracy-oriented SBRSs, called DCA-SBRS, to help them generate a more diversified recommendation list without significantly sacrificing their accuracy performance. Given the widespread adoption and efficacy of attention mechanisms in existing state-of-the-art accuracy-oriented SBRSs \cite{wang2021survey,under2023}, we extend our approach by incorporating category information into the attention mechanism. Specifically, DCA-SBRS is composed of two particularly designed parts: (1) a Model-agnostic Diversity-oriented Loss (MDL) function, working with accuracy-oriented loss (e.g., cross-entropy loss), exploits items' category attribute and estimated item scores from the given SBRS; and (2) a Non-invasive Category-aware Attention (NCA) mechanism, which inspired by NOVA~\cite{liu2021noninvasive} utilizes category information in a non-invasive way, instead of directly fusing category information, and acts as directional guidance (attention signal) to help more accurate session-based recommendation. 
The main contributions of this work are summarized as follows:
\begin{itemize}
    \item We propose a simple yet effective diversity-oriented loss function that can be used as a 
    model-agnostic and individual plugin to deep neural accuracy-oriented SBRSs to improve their diversity performance, mitigating the technical gap between accuracy-oriented and diversified SBRSs.
    \item 
    We transfer the non-invasive idea from NOVA~\cite{liu2021noninvasive} into the common attention mechanism used in SOTA accuracy-oriented SBRSs (e.g., NARM and GCE-GNN) to capture more accurate preference by utilizing category information in a non-invasive way, so as to efficiently help maintain recommendation accuracy.
    \item We conduct extensive experiments on three real-world datasets, in terms of accuracy, diversity, and comprehensive performance (jointly considering accuracy and diversity), to demonstrate the effectiveness of our DCA-SBRS framework. Experimental results unveil that,
    our framework can help SOTA SBRSs achieve extraordinary performance in terms of diversity and comprehensive performance 
    {(e.g., average 74.1\% and 52.3\% increase on ILD$@$10 and F-score$@$10 respectively)}, without significantly deteriorating recommendation accuracy in contrast with SOTA diversified SBRSs 
    {(e.g., an average of only $1.6\%$ decrease on accuracy regarding NDCG@$10$ but $138\%$ increase on diversity for ILD@$10$ on Diginetica)}. Additionally, we fairly analyze the limitations of the standard comprehensive measure and offer alternative solutions.
\end{itemize}

\section{Related work}
Our study is related to two major areas: session-based recommendation, and diversified recommendation.

\vspace{-2mm}
\subsection{Session-Based Recommendation}\label{subsec:sbr}

The approaches on SBRSs can be divided into two groups: conventional non-neural methods and deep neural ones.
Typical conventional techniques include but are not limited to Item-KNN \cite{sarwar2001item}, BPR-MF~\cite{rendle2009bpr}, and FPMC~\cite{rendle2010factorizing}. For example, FPMC deploys Matrix Factorization (MF) with Markov Chain (MC) to better deal with dependent relationships between items in sequence. However, they generally suffer from inadequately addressing the item relationships in comparatively longer sequences.
In contrast, deep neural networks can better deal with much longer sequences and thus generate more effective recommendation \cite{tan2016improved,hidasi2018recurrent}.
For example, GRU4Rec~\cite{hidasi2015session} and its variants~\cite{tan2016improved,hidasi2018recurrent} apply GRU to capture the long-term dependency in a sequence.
NARM~\cite{li2017neural} further adopts an attention mechanism to assess the similarity between previous items and the last item in every session, and the hidden states are then weighted averaged to obtain the main-purpose session representation. And, STAMP~\cite{LiuZMZ18} models both users’ general interests and current interests using attentive nets and basic multiple-layer perceptions (MLPs) instead of adopting RNNs.

However, the above techniques only model one-way transitions between successive items, ignoring transitions between contexts (i.e., other items in the session)~\cite{qiu2019rethinking}. Recently, GNNs have been employed to mitigate the research gap \cite{yu2020tagnn}. For instance, SR-GNN~\cite{WuT0WXT19} and GC-SAN~\cite{XuZLSXZFZ19} import GNNs to generate more accurate item embedding vectors based on the current session graph built for each session. Besides the current session graph, GCE-GNN~\cite{wang2020global} also explores item relationships in the global session graph.

It is worth noting that, the above conventional and deep neural SBRSs are all accuracy-oriented approaches that fail to consider diversity (i.e., non-diversified). Given that RSs have an iterative or closed feedback loop, this may result in filter bubbles~\cite{nguyen2014exploring,khenissi2020theoretical}.

\vspace{-3mm}
\subsection{Diversified Recommendation}\label{subsec:dr}
Towards individual diversity in traditional RSs, inspired by dissimilarity score in Maximal Marginal Relevance (MMR) \cite{carbonell1998use}, some studies \cite{agrawal2009diversifying,santos2010exploiting} define diversification on explicit aspects (categories) or sub-queries.
Besides, DPP is utilized \cite{chen_fast_2018,kulesza2012determinantal,wu2019pd,gan2020enhancing} to provide a better relevance-diversity trade-off in recommendation as it can score sets of items collectively and consider negative correlations between various items. The aforementioned studies are two-stage ones which re-rank items accounting for diversity in the second stage. In traditional RS, there are only several end-to-end studies~\cite{zheng2021dgcn,liang2021enhancing} which simultaneously optimize diversity and accuracy.

\begin{figure}[t]
    \centering
    \includegraphics[scale=1.2]{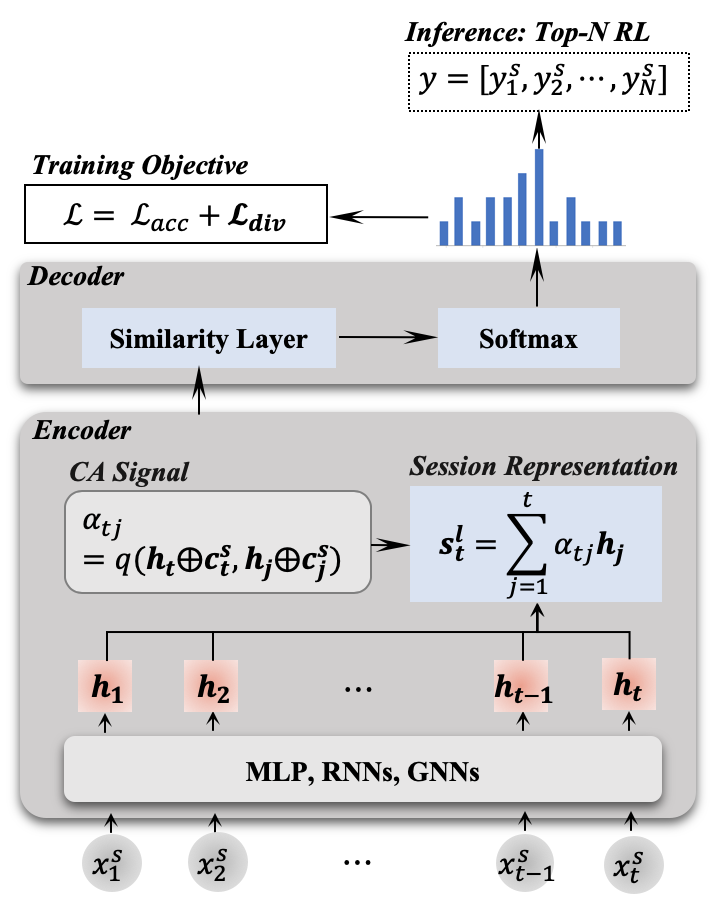}
    \caption{An Overview of Our Proposed DCA-SBRS.}
    \label{fig:overall}
    \vspace{-0.1in}
\end{figure}

To the best of our knowledge, there are only three diversified (and also end-to-end) works for session-based recommendation: MCPRN \cite{Wang0WSOC19}, ComiRec~\cite{Cen2020ControllableMF}, and IDSR~\cite{chen2020improving}.
Specifically, MCPRN uses mixture-channel purpose routing networks to guide multi-purpose learning, while ComiRec explores two methods as multi-interest extraction modules(i.e., the dynamic routing and self-attentive methods). Thus, multiple session representations are used by MCPRN and ComiRec to capture user preferences which can implicitly satisfy user needs.
In contrast, IDSR delivers the end-to-end recommendation under the guidance of the intent-aware diversity promoting (IDP) loss and explicitly creates set diversity. A ``trade-off hyper-parameter" (in IDSR) is adopted to keep the balance between recommendation relevance and diversity.

To summarize, such diversified designs in those three works cannot be easily adapted to existing representative accuracy-oriented SBRSs. Besides, regarding the widely-hold ``trade-off" relationship, these studies fail to obtain a satisfying performance on recommendation accuracy (can also be observed in Tables \ref{tab:digi}-\ref{tab:tmall}).

\section{Our DCA-SBRS Framework}

In this section, we firstly formulate our research problem, and then introduce the two components in the proposed framework in detail.

\vspace{-2mm}
\subsection{Problem Statement and Model Overview}
Let $\mathcal{X}=\{x_1, x_2, \cdots, x_m\}$ be all of items and $\mathcal{C}=\{c_1, c_2, \cdots, c_n\}$ be all of categories.
Each anonymous session, denoted by $S=[x_1^s, x_2^s,\cdots, x_t^s]$, consists of item IDs in chronological order (i.e., items clicked by a user), where $x_i^s$ denotes the $i$-th item clicked within session $S$. Additionally, our framework uses the category attribute of items (i.e., $c_i^s$ denotes the corresponding category of $x_i^s$) to guide the session representation learning for better item prediction.
\emph{Given a session $S$, the objective of our session-based recommendation aims to recommend a both diversified and accurate Top-$N$ item list, denoted as $y=[y_1^s, y_2^s, \cdots, y_N^s]$, for next-item prediction.}

To address the problem, we propose a Diversified Category-aware Attentive framework which can be instantiated with SOTA accuracy-oriented SBRS, named DCA-SBRS, to improve the diversity performance of the corresponding SBRS while preserving its recommendation accuracy. It mainly consists of two novel components: 1)
{Model-agnostic Diversity-oriented Loss function}
(\textbf{MDL}, $L_{div}$), working with accuracy-oriented loss (e.g., cross-entropy loss $L_{acc}$), which is built on items' category attribute and estimated item scores by the SBRS. It can help achieve more diverse recommendation lists towards existing SOTA accuracy-oriented SBRSs; 2) {Non-invasive Category-aware Attention}
(\textbf{NCA}) mechanism, which utilizes category information as directional guidance to replace normal attention mechanism widely used in existing SBRSs. With such design, since there exists a widely-known ``trade-off" relationship between recommendation accuracy and diversity \cite{chen2020improving}, the adverse effect induced by diversity objective on recommendation accuracy can be partially alleviated.

Figure~\ref{fig:overall} presents the architecture of our DCA-SBRS framework, which depicts the installation of the MDL and NCA components on the basis of the general encoder-decoder framework and common attention mechanism from a SOTA SBRS, NARM~\cite{li2017neural}. Without losing generality, as shown in Figure \ref{fig:overall}, let encoder-decoder framework denotes the architecture of SOTA SBRSs where the encoder is to encode session representation, while the decoder is designed to estimate item scores for generating recommendations. The 
{similarity layer} projects the session representation into the item space, and then produces a Top-$N$ recommendation list. We next present the two components in detail.

\vspace{-2mm}
\subsection{Model-agnostic Diversity-oriented Loss}

\begin{figure}[ht]
    \centering
    \includegraphics[width=.4\textwidth]{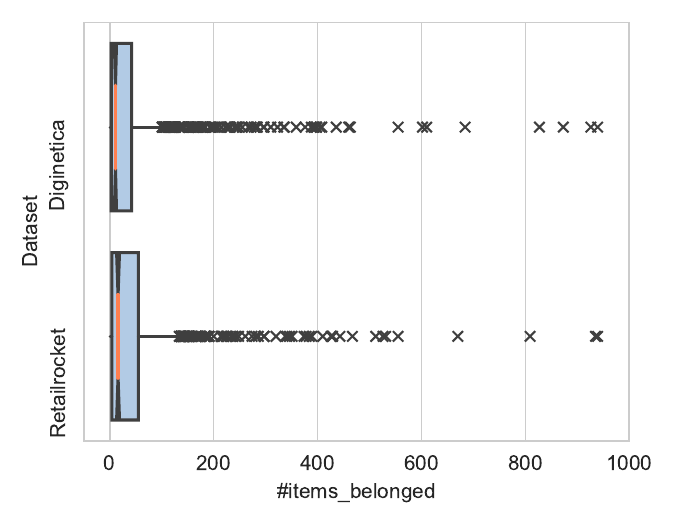}
    \vspace{-0.15in}
    \caption{The Unbalanced Grouping Induced by the Category (the symbol `$\times$' denotes the outliers with a mass of involved items).}
    \label{fig:unbalanced}
    \vspace{-0.1in}
\end{figure}

{The goal of this module is to enhance diversity performance by acting as a model-agnostic plugin to accuracy-oriented SBRSs. The non-diversified SBRSs frequently predict relevance scores of items by capturing preferences from item sequences. For simplicity, we attempt to leverage the obtained relevance scores as the foundation of this module and increase recommendation diversity by penalizing more monotonous \underline{R}ecommendation \underline{L}ist (e.g., most items in a top-N recommended list of the same category).}
To fulfill the goal, as shown in Figure \ref{fig:overall}, the model-agnostic diversity-oriented loss ($L_{div}$) is designed to facilitate existing SBRSs achieve the end-to-end learning. 
Specifically, we define it via using the entropy of estimated category distribution $\widehat{P}_c$ 
{in a recommended list}, given by,
\begin{equation}\label{equ:l_div}
    L_{div}=-\mathrm{H}(\widehat{P}_c),
\end{equation}
where $\mathrm{H}(\widehat{P})=-\sum_j \widehat{P}_{c_j} log_2 \widehat{P}_{c_j}$ (${c_j}\in\mathcal{C}$) denotes the information entropy.
{A larger $\mathrm{H}(\widehat{P}_c)$ depicts that the recommended list is likely to be more diverse from the category perspective. In this case, its negative value can be regarded as penalizing the recommended list with low diversity.}
Intuitively, the reasonable $\widehat{P}_{c_i}$ ($c_i\in\mathcal{C})$ in a recommendation list should satisfy the following two characteristics:
\begin{itemize}
    \item \textbf{\emph{In proportion to the number of items from the category $c_i$}}: In real-world datasets, the grouping induced by categorical attribute can be very unbalanced~\cite{zhao2018categorical}. For better understanding, we select two datasets (Diginetica and Retailrocket) and statistically show the number of items belonging to the same category using Box-plot as Figure~\ref{fig:unbalanced}. As can be observed, the outliers in the Box-plot depict that for some categories, a large group of items are involved while for others only a few. The category with a larger group of items is more likely to appear in the RL without considering \emph{personalized preference}.
    \item \textbf{\emph{In proportion to relevance scores of items}}: Regarding personalized preference, representative SBRSs recommend Top-$N$ items by ranking the predicted scores given session $S$. As a result, the items with much higher scores are more likely to appear in the RL along with their corresponding categories.
\end{itemize}

Considering that common accuracy-oriented SBRSs only output predicted item scores without a special module capturing category scores, we simulate the category distribution in the 
RL,
which can well satisfy the above two characteristics as below,
\begin{equation}\label{eq:cd}
\widehat{P}_{c_i}=\sum_{c(x_j)=c_i}\widehat{P}_{x_j},
\end{equation}
where $\widehat{P}_{x_j}$ depicts the predicted personalized preference score of item $x_j$ obtained by the given SOTA SBRS ($\sum\widehat{P}_{x_j}=1$ using softmax function on all items). We sum the scores of items from the category $c_i$ as the occurred probability of category $c_i$ so as to consider both the number of items in $c_i$ and personalized preference $\widehat{P}_{x_j}$.
Then, $L_{div}$ combined with the origin accuracy-oriented loss $L_{acc}$ (e.g., the cross-entropy of the prediction results~\cite{li2017neural,wang2020global}) is the final loss function for model training,
\begin{equation}\label{equ:final-loss}
    L = L_{acc} + \lambda L_{div},
\end{equation}
where $\lambda$ controls the importance of our proposed MDL.

\begin{table*}[t]
\centering
\caption{The Category-aware Attentive Signal Extension of 
Representative SBRSs (the symbols used in these functions are aligned with the ones used in the original papers).}
\vspace{-0.1in}
\begin{adjustbox}{max width=0.8\linewidth}
\begin{tabular}{l|l|l}\toprule
                         & Attention Signal  & Category-aware Attention Signal   \\\midrule
\multirow{2}{*}{NARM}    & $\alpha_{tj} = \bf{v}^T\sigma (\bf{A_1} \bf{h_t} + \bf{A_2} \bf{h_j})$
                         & $ \alpha_{tj} = \bf{v}^T\sigma (\bf{A_1} (\bf{h_t}+\bf{c_t^s}) + \bf{A_2} (\bf{h_j}+\bf{c_j^s}))$ \\
                         & $ \bf{c_t^l}=\sum_{j=1}^t \alpha_{tj} \bf{h_j}$                 
                         &  $ \bf{c_t^l}=\sum_{j=1}^t \alpha_{tj} \bf{h_j}$                \\\hline
\multirow{3}{*}{STAMP}   & $ \bf{m_s}=\frac{1}{t}\sum_{i=1}^t \bf{x_i}$ & $ \bf{m_s}=\frac{1}{t}\sum_{i=1}^t (\bf{x_i}+\bf{c_i})$ \\
                         &  $ \alpha_i = \bf{W_0}\sigma\left(\bf{W_1}\bf{x_i}+\bf{W_2}\bf{x_t}+\bf{W_3}\bf{m_s}+\bf{b_a}\right)$   
                         &  $ \alpha_i = \bf{W_0}\sigma\left( \bf{W_1}(\bf{x_i}+\bf{c_i})+\bf{W_2}(\bf{x_t}+\bf{c_t})+\bf{W_3}\bf{m_s}+\bf{b_a}\right)$              \\
                         & $ \bf{m_a}=\sum_{i=1}^t \alpha_i \bf{x_i}$                  
                         & $ \bf{m_a}=\sum_{i=1}^t \alpha_i \bf{x_i}$          \\\hline
\multirow{3}{*}{GCE-GNN} & $ \bf{z}_i=\tanh \left(\bf{W}_3\left[\bf{h}_{v_i^s}^{\prime} \| \bf{p}_{l-i+1}\right]+\bf{b}_3\right)$
                         &  $ \bf{z}_i=\tanh \left(\bf{W}_3\left[\bf{h}_{v_i^s}^{\prime} \| \bf{p}_{l-i+1} \| \bf{c}_i^s\right]+\bf{b}_3\right)$\\
                         & $ \bf{s}^{\prime}=\frac{1}{l} \sum_{i=1}^l \bf{h}_{v_i^s}^{\prime}$                 
                         & $ \bf{s}^{\prime}=\frac{1}{l} \sum_{i=1}^l \left(\bf{h}_{v_i^s}^{\prime} + \bf{c}_l^s\right)$                 \\
                         & $ \beta_i=\bf{q}_2^{\top} \sigma\left(\bf{W}_4 \bf{z}_i+\bf{W}_5 \bf{s}^{\prime}+\bf{b}_4\right)$                  
                         &  $ \beta_i=\bf{q}_2^{\top} \sigma\left(\bf{W}_4 \bf{z}_i+\bf{W}_5 \bf{s}^{\prime}+\bf{b}_4\right)$                \\                         
                         & $ \bf{S}=\sum_{i=1}^l \beta_i \bf{h}_{v_i^s}^{\prime}$                  
                         & $ \bf{S}=\sum_{i=1}^l \beta_i \bf{h}_{v_i^s}^{\prime}$                \\\bottomrule
\end{tabular}
\end{adjustbox}
\label{tab:ca_details}
\end{table*}

\subsection{Non-invasive Category-aware Attention}
There exists a widely-known ``trade-off" relationship between recommendation accuracy and diversity \cite{chen2020improving}. In this case, the plugged diversity loss (in MDL) will probably lead to deteriorating performance on recommendation accuracy towards accuracy-oriented SBRSs. To address this issue, we consider to exploit category information to enhance preference learning.
 
As shown in Figure~\ref{fig:mtl_compar}, invasive fusion (like merely concatenating item embeddings with the relevant category embeddings as input), might not considerably improve recommendation accuracy.
Therefore, 
considering that attention mechanisms are widely adopted by SOTA accuracy-oriented SBRSs, we transfer the non-invasive idea from NOVA~\cite{liu2021noninvasive} into the common attention mechanism.
Specifically, 
as shown in Figure~\ref{fig:overall}, the encoder, employing a deep learning technique as an existing SBRS (e.g., RNN~\cite{li2017neural}, MLP~\cite{LiuZMZ18}, or GNN~\cite{WuT0WXT19,wang2020global}), firstly coverts session $S=[x_1^s, x_2^s,\cdots, x_t^s]$ into a set of high-dimensional hidden states $\bf{h}=[\bf{h_1}, \bf{h_2},\cdots, \bf{h_t}]$,
which are weighted summed by attention signal output by common attention mechanism at time $t$ (denoted as $\bf{\alpha}_t=\{\alpha_{t1},\ldots,\alpha_{tt}\}$) to obtain the current session representation decoded at time $t$ (denoted as $\bf{s_t}$).

{The category-aware extensions for SOTA SBRSs with attention mechanism (i.e. NARM, STAMP, and GCE-GNN) are described in detail in Table~\ref{tab:ca_details}, where the symbols in the functions are unified with the original papers and thus the corresponding detailed explanation is omitted here. Note that $\bf{c_j^s}$ is the corresponding category embedding vector of item $x_j^s$ in session $S=[x_1^s, x_2^s,\cdots, x_t^s]$.}

Here, we use NARM~\cite{li2017neural} as an example to further elaborate our 
{NCA}. In NARM, the attention signal $\alpha_{tj}$ is computed as the correlation between the final hidden state $\bf{h_t}$ and the hidden state of the $j$-th item, $\bf{h_j}$,
\begin{equation}
    \alpha_{tj} = q(\bf{h_t}, \bf{h_j}) = \bf{v}^T\sigma (\bf{A_1} \bf{h_t} + \bf{A_2} \bf{h_j}),
\end{equation}
where $\sigma$ is an activate function (e.g., sigmoid function) and matrix $\bf{A_1}, \bf{A_2}$ are used to transform hidden states into a latent space, respectively. Correspondingly, our NCA mechanism further uses the category attribute as directional guidance and keeps the hidden states undoped in their vector space. Specifically, 
{NCA} uses the category attribute to update the attention signal as:
\begin{equation}
\begin{aligned}
 \alpha_{tj} &= q(\bf{h_t}\oplus \bf{c_t^s}, \bf{h_j}\oplus \bf{c_j^s}) \\
 &= \bf{v}^T\sigma (\bf{A_1} (\bf{h_t}+\bf{c_t^s}) + \bf{A_2} (\bf{h_j}+\bf{c_j^s})),   
\end{aligned}
\end{equation}
where $\bf{c_j^s}$ is the corresponding category embedding vector of item $x_j^s$ and $\oplus$ denotes element-wise addition.
Note that here we use the simplest fusor `addition' to straightforwardly add the hidden states and category embedding vectors in this paper. It can also be replaced by other fusors, like `concatenation' or `gating'~\cite{liu2021noninvasive}.

To conclude, by doing this, we have successfully exploited category information in a non-invasive way 
to help generate attention signals, with the goal of maintaining the recommendation accuracy.  

\subsection{Discussion: Simple yet Effective Approach}\label{subsec:simple}
Our DCA-SBRS framework can serve as a plugin for SOTA
accuracy-oriented SBRSs to improve their diversity performance with MDL module while in the meantime striving to maintain their
recommendation accuracy with NCA mechanism.
Generally speaking, 
both MDL module and NCA mechanism can be easily equipped with existing SOTA accuracy-oriented SBRSs to further promote their performance regarding diversity towards more trustworthy recommender systems \cite{ge2022survey,wang2022trustworthy}. Extensive experimental results in Section \ref{sec:results} verify that our approach can help SOTA SBRSs (i.e., NARM, STAMP, and GCE-GNN) obtain extraordinary performance in terms of recommendation diversity and comprehensive performance (considering both accuracy and diversity).

Besides, our approach is much more lightweight (simple yet effective) than existing diversified recommender systems: (1)
{in contrast to the (two-stage) re-ranking methods (e.g., MMR~\cite{carbonell1998use}), MDL can achieve end-to-end learning, that is, simultaneously maximizing accuracy and diversity objectives; (2) unlike other diversified SBRSs (e.g., IDSR~\cite{chen2020improving}) relying on specifically calibrated diversity-aware components with a substantial amount of extra parameters, our MDL module is a model-agnostic plugin by utilizing the estimated relevance scores of items from every existing SOTA SBRS and the category information, which thus requires limited extra parameters and is efficiently comparable to the corresponding SBRS; and (3) both MMR~\cite{carbonell1998use} and IDSR~\cite{chen2020improving} employ a greedy iterative inference algorithm to generate the final Top-$N$ recommended lists. On the contrary, our DCA-SBRS framework directly generate a recommended list including Top-N items with the highest final scores, implying that our approach is more computationally efficient in model inferences. It is also empirically verified in table~\ref{tab:time_complexity}.}


\begin{table}[t]
    \centering
    \addtolength{\tabcolsep}{4pt}
    \caption{Statistics of Datasets (Note: \# train and \#test are the number of sessions before sequence splitting preprocess; avg. len. denotes the average session length; DS is the diversity score defined in Section~\ref{subsec:metrics}; and RR~\cite{repeat2019} is the repeat ratio, indicating the ratio of repeated items within a session.).
    }
    \begin{tabular}{lrrrr}
    \toprule
    Dataset &  Diginetica & Retailrocket & Tmall\\
    \midrule
    \# interactions & 993,483 & 1,040,796 & 1,505,683\\
    \# train & 186,670 & 283,446 & 188,756 \\
    \# test &  18,101 & 11,718 & 51,894 \\
    \# items &  43,097 & 45,831 & 96,182 \\
    \# categories & 995 & 871 &  822 \\
    avg. len. & 4.8504 & 3.5262 & 6.0775 \\
    \hline\hline
    train DS & 0.3741  & 0.4646 & 0.6575 \\
    test DS & 0.3721 & 0.4893 &  0.6278 \\
    train RR & 0.1301 & 0.2488 &  0 \\
    test RR & 0.1317 & 0.2370 & 0 \\
    \bottomrule
    \end{tabular}
    \label{tab:updated_datasetStatistics}
    \vspace{-0.1in}
\end{table}

\section{Experimental Settings}
In this section, we introduce the selection of datasets, baselines, and evaluation metrics. The specifics of dataset preprocessing and partitioning, as well as the hyper-parameter settings for our methods and other baselines, are also provided.
The source code and datasets are available online\footnote{\url{https://github.com/qyin863/DCA-SBRS}.}.

\subsection{Datasets and Preprocessing}\label{subsec:datasets}
For the experimental purpose, we delicately select three representative public e-commerce datasets (i.e., Diginetica\footnote{\url{https://competitions.codalab.org/competitions/11161\#learn_the_details-overview}.}, Retailrocket\footnote{\url{https://www.kaggle.com/retailrocket/ecommerce-dataset}.}, Tmall\footnote{\url{https://tianchi.aliyun.com/dataset/dataDetail?dataId=42}.}) with item category information, following~\cite{wang2021survey,li2017neural,wang2020global}.

\begin{itemize}
    \item \textbf{Diginetica} from CIKM Cup 2016, contains user sessions, taken from records of an e-commerce search engine with its own `SessionId'. We only use the data with the behavior type `view'.
    \item \textbf{Retailrocket} collects users' interactions on an e-commerce website over a period of 4.5 months. We select interactions with the behavior type `view', and a new session is created when the user's idle time exceeds 30 minutes following~\cite{luo2020collaborative}.
    \item \textbf{Tmall} from the IJCAI-15 competition, includes anonymous Tmall shopping logs. We adopt interactions with the behavior type `buy' and `view', and partition user history into sessions by day following~\cite{ludewig2018evaluation}.
    We pick $1/16$ sessions as a sampling inspired by Yoochoose fractions~\cite{li2017neural}.
\end{itemize}

For data preprocessing, following~\cite{li2017neural,LiuZMZ18,WuT0WXT19}, we filter out sessions of length $1$ and items occuring less than $5$ times. Then  we set the most recent data (i.e., the last one week) as the test set and the previous sessions as the training set. The validation set contains the final week of data from the training set. Additionally, we drop items appearing in the test set but not in the training set. The statistics of these three datasets after preprocessing are shown in Table~\ref{tab:updated_datasetStatistics}.
A sequence splitting preprocess, 
that is, generating $n-1$ sub-sequences $([i_1], i_2)$, $([i_1, i_2], i_3)$, $\dots$, $([i_1,\dots, i_{n-1}], i_n)$ for a session sequence $S=[i_1, i_2, \dots, i_n]$, is required if a recommendation model is not trained in session-parallel manner \cite{hidasi2015session}.

\subsection{Baseline Models}\label{subsec:baselines}
To explore the recommendation performance on accuracy and diversity, following~\cite{wang2021survey,wang2020global,chen2020improving}, we select three categories of popular and representative baseline models for \underline{session-based recommendation}, including \emph{traditional methods}, \emph{deep neural methods with attention mechanism} (as they are chosen as the basic predictors in our proposed framework), and \emph{deep diversified methods}.

\smallskip\noindent\textit{1. Traditional Methods.}
\begin{itemize}
    \item \textbf{Item-KNN}~\cite{sarwar2001item}
    measures cosine similarity of every two items regarding sessions in the training data. It recommends items for a session that are most similar to the last item.
    \item \textbf{BPR-MF}~\cite{rendle2009bpr} 
    performs Matrix Factorization (MF) with a pairwise ranking loss.
    Particularly, the session feature vector is averaged over all items in the session.
\end{itemize}

\smallskip\noindent\textit{2. Deep Neural Methods with Attention Mechanism.}
\begin{itemize}

    \item \textbf{NARM}~\cite{li2017neural} is an RNN-based model with an attention mechanism, which combines the last hidden vector and the main purpose from the hidden states as the final representation to produce recommendations.
    
    \item \textbf{STAMP}~\cite{LiuZMZ18} applies attention layers on item representations directly and captures the user’s long-term preference as well as short-term interest from the session context.
    

    \item \textbf{GCE-GNN}~\cite{wang2020global} constructs both the local (current session) and global (all sessions) graphs to obtain session- and global-level item embeddings. Then, before the soft attention, it incorporates the reversed position information into the item embedding.

\end{itemize}

\smallskip\noindent\textit{3. Deep Diversified Methods.}

\begin{itemize}
    \item \textbf{MCPRN}~\cite{Wang0WSOC19} models users’ multiple purposes of the session, rather than only one purpose 
    in common SBRSs. Furthermore, it combines the above various learned purposes by the target-aware attention to get the final representation. As stated in the original paper, MCPRN can boost both accuracy and diversity.
        \item \textbf{NARM+MMR}~\cite{chen2020improving} is a two-stage approach which in the second stage uses MMR~\cite{carbonell1998use} and a greedy algorithm to re-rank items provided by NARM in terms of relevance scores in the first stage. 
    \item \textbf{IDSR}~\cite{chen2020improving} is the first end-to-end deep neural network for SBRSs that takes both diversity and accuracy into account. The hyper-parameter $\lambda$ is used to balance the relevance score and diversification score.
\end{itemize}

\subsection{Evaluation Metrics}\label{subsec:metrics}
We adopt the following metrics related to accuracy, diversity, and both to conduct a thorough evaluation. Higher metric values indicate better performance. 
%
Towards \textit{accuracy}, we select HR (Hit Rate), MRR (Mean Reciprocal Rank), and NDCG (Normalized Discounted Cumulative Gain) by following state-of-the-arts~\cite{li2017neural,WuT0WXT19,wang2020global}. Specifically, \textbf{HR} depicts whether the Top-$N$ Recommended List (abbreviated as RL, and $N$ is the length of the RL) contains the target item;
\textbf{MRR} and \textbf{NDCG} both measure the hit position and encourage the predicted item to rank ahead in the recommended list.
%
Towards \textit{diversity}, 
we choose the widely-used ILD (Intra-List Distance)~\cite{Cen2020ControllableMF,chen2020improving}, 
Entropy~\cite{Wang0WSOC19,zheng2021dgcn}, and Diversity Score~\cite{liang2021enhancing} as the evaluation metrics. Particularly, \textbf{ILD} measures the average distance between each pair of items in the recommended list,

\begin{equation}
        \text{ILD} = \frac{\sum_{(i,j)\in RL}d_{ij}}{|RL|\times(|RL|-1)},\label{eq:ild}
\end{equation}
where $d_{ij}$ represents the euclidean distance between the respective embeddings (e.g., one-hot encoding) of categories that items $i$ and $j$ belong to.

\textbf{Entropy} measures the entropy of item category distribution in the recommended list;
and \textbf{Diversity Score} (shorted as \textbf{DS}) is calculated by the number of interacted/recommended categories divided by number of interacted/recommended items.
Additionally, we use \textbf{F-score}~\cite{hu2017diversifying}, the harmonic mean of HR and ILD, as an aggregative indicator capturing both accuracy and diversity.
%
%

\subsection{Hyper-parameter Settings}

For a fair comparison, we use the Bayesian TPE\footnote{Compared to the grid and random search, it has proven to be a more intelligent and effective technique, especially for deep methods (having more hyper-parameters)~\cite{sun2020we}.}~\cite{bergstra2011algorithms} of Hyperopt\footnote{\url{https://github.com/hyperopt/hyperopt}} framework to tune hyper-parameters of all methods according to their performance 
on the validation set (i.e., the last week of the training set).
We have integrated all the codes with PyTorch framework, except for IDSR. Specifically, we adopt its official code\footnote{\url{https://bitbucket.org/WanyuChen/idsr/}} with its own early-stopping mechanism.
For all methods, Adam is utilized as the model optimizer; the dimension of item embedding is searched in the range of $[100, 300]$ stepped by 50; the learning rate is searched in $\{0.001,0.005,0.01,0.05\}$; the size of mini-batch is searched from $\{64,128,256,512\}$; the number of epochs is searched in the range of $[10, 40]$ stepped by 5.  
The exceptions are made on GCE-GNN, where we set its dimension of item embedding and size of mini-batch as $100$ (consistent with the original paper setting) due to memory space limitations; and set the size of mini-batch as $50$ for MCPRN. 
For IDSR, we search $\lambda$, which balances the importance of relevance and diversification scores, in  $\{0.2,0.5,0.8\}$ on every dataset. 
Moreover, for NARM+MMR, we set the multiplier $\lambda=5e-6$ for the diversification score in MMR, so as to avoid a significant decrease (e.g., more than 20\% decline) on accuracy performance in comparison with NARM.
%
The detailed best hyper-parameter settings are shown in Table~\ref{tab:optimalhypers}.

\begin{table*}[htbp]
\centering
\footnotesize
\caption{The Optimal Hyper-parameter Settings by Bayesian TPE of Hyperopt.}
\vspace{-0.1in}
\begin{adjustbox}{max width=\textwidth}
\begin{tabular}{l|l|c|c|c|l|l}
\toprule
Model                   & Hyper-parameter       &  Digi* & Retail* & Tmall & Searching Space             & Description \\\midrule
Item-KNN                & -alpha                &0.9270            & 0.7100           & 0.8514             & $\mathcal{U}(0.1,1)$           &Balance for normalizing items' supports             \\\hline
\multirow{4}{*}{BPR-MF} & -item\_*\_dim &300            & 100           &200                & $[min=100,max=300,step=50]$       & the dimension of item embedding             \\
                        & -lr                   &0.01            & 0.01           &0.001                  & $[0.001,0.005,0.01,0.05]$ & learning rate             \\
                        & -batch\_size          &64           &64            &512                  & $[64,128,256,512]$        & the size for mini-batch            \\
                        & -epochs               &20            & 20           &40              & $[min=10,max=40,step=5]$         &the number of epochs             \\\hline 

\multirow{6}{*}{NARM} & -item\_*\_dim &200            &100            &250       & $[min=100,max=300,step=50]$       &            \\
                        & -lr                   &0.001            &0.001            &0.005      & $[0.001,0.005,0.01,0.05]$ &           \\
                        & -batch\_size          &512           &512            &256       & $[64,128,256,512]$        &         \\
                        & -epochs               &35            &40            &25     & $[min=10,max=40,step=5]$         &        \\
                        & -hidden\_size               &50            &150            &150    & $[min=50,max=200,step=50]$          &   the dimension of latent vector          \\
                        & -n\_layers               &1            &1            &1    & $[1,2,3]$         &   the number of layers in RNN          \\\hline 
\multirow{6}{*}{DCA-NARM} & -item\_*\_dim &100 & 100 &200 &&      \\
                        & -lr                   &0.001 & 0.001 &0.005 &&         \\
                        & -batch\_size        &512 &512 &256 &&          \\
                        & -epochs               &20 &15 &20 &&          \\
                        & -hidden\_size             &200 &100 &150 &&          \\
                        & -n\_layers             &1 &2 &1 &&     \\\hline 
\multirow{4}{*}{STAMP} & -item\_*\_dim &100            &100            &150   & $[min=100,max=300,step=50]$       &             \\
                        & -lr                   &0.001            &0.001            &0.01     & $[0.001,0.005,0.01,0.05]$ &             \\
                        & -batch\_size          &128            &512            &256  & $[64,128,256,512]$        &             \\
                        & -epochs               &35            &20            &35   & $[min=10,max=40,step=5]$         &             \\\hline     
\multirow{4}{*}{DCA-STAMP} & -item\_*\_dim &100 &200 &200 && \\
                        & -lr    &0.001 &0.001 &0.01 &&    \\
                        & -batch\_size   &256 &512 &512 && \\
                        & -epochs     &35 &15 &40 &&   \\\hline  
\multirow{7}{*}{GCE-GNN} & -item\_*\_dim &250            &100            &100     & $[100]$        &             \\
                        & -lr                   &0.001            &0.001            &0.005        & $[0.001,0.005]$ &             \\
                        & -batch\_size          &128           &100            &100   & $[100]$        &             \\
                        & -epochs               &10            &30            &20    & $[min=10,max=30,step=5]$          &             \\
                        & -n\_iter               &1            &1            &2      & $[1,2]$          & the number of hop             \\
                        & -dropout\_gcn               &0.4            &0.4            &0.2     & $[0,0.2,0.4,0.6,0.8]$          &dropout rate             \\                
                        & -dropout\_local               &0.5            &0.0            &0.0      & $[0,0.5]$         &dropout rate             \\\hline 
\multirow{7}{*}{DCA-GCEGNN} & -item\_*\_dim &100 &100 &100 &&    \\
                        & -lr      &0.001 &0.001 &0.005 &&  \\
                        & -batch\_size     &100 &100 &100 &&\\
                        & -epochs     &30 &20 &20 &&   \\
                        & -n\_iter       &2 &2 &2 &&  \\
                        & -dropout\_gcn      &0.0 &0.2 &0.4 &&   \\                
                        & -dropout\_local      &0.0 &0.0 &0.0 &&    \\\hline 
                        
\multirow{6}{*}{MCPRN} & -item\_*\_dim &150            &150            &100       & $[min=100,max=200,step=50]$        & dimension of item embedding/latent vector              \\
                        & -lr                   &0.005            &0.005            &0.005        & $[0.005,0.01,0.05]$ &             \\
                        & -batch\_size          &256            &50            &50     & $[50]$        &             \\
                        & -epochs               &15            &30            &25     & $[min=10,max=40,step=5]$         &             \\        
                        & -tau               &1            &0.01            &0.01         & $[0.01,0.1,1,10]$          & temperature parameter in softmax             \\ 
                        & -purposes               &1            &4            &1     & $[1,2,3,4]$          &The number of channels             \\ \hline
\multirow{6}{*}{Remark}  & \multicolumn{6}{l}{1. Digi* represents Diginetica, Retail* for Retailrocket, item\_*\_dim for item\_embedding\_dim.}\\
                         & \multicolumn{6}{l}{2. Omit the hyper-parameter description if exists before.}\\
                         & \multicolumn{6}{l}{Additionally, DCA-SBRS and the related SBRS have the same searching space, hence omit.}\\
                         & \multicolumn{6}{l}{3. Due to memory limit, set item\_*\_dim, batch\_size as 100 (original setting) in GCE-GNN, and batch\_size as 50 in MCPRN except Digi*.}\\
                         & \multicolumn{6}{l}{4. IDSR uses own official TensorFlow code with early-stopping. Tune $\lambda_e \in [0.1,1]$ and set it as 1 for four datasets. Besides, tune the}\\
                         & \multicolumn{6}{l}{trade-off hyper-parameter $\lambda$ from $\{0.2,0.5,0.8\}$ aiming competitive accuracy and set it as $0.8,0.5,0.8$ for three datasets respectively.}\\
                        \bottomrule
\end{tabular}
\end{adjustbox}
\label{tab:optimalhypers}
\end{table*}

\section{Experimental Results}\label{sec:results}
In this section, we evaluate the performance of DCA-SBRS on the three selected real-world datasets to verify its superiority (in comparison with other SOTA methods) and the effectiveness of its respective modules. 
Additionally, we analyze the shortcomings of the standard comprehensive measurement to measure both accuracy and diversity (i.e., F-score), and provide remedies accordingly.


\subsection{Overall Comparisons}
Tables~\ref{tab:digi}-\ref{tab:tmall} exhibit the experimental results of the chosen baselines on the three real-world datasets, where the best result for each 
metric is highlighted in boldface and the runner-up is underlined; the row `Improvements' indicates the average relative enhancements achieved by our DCA-SBRSs over the corresponding SBRSs on various metrics across the three datasets, as shown in Equation~\ref{eq:improv}. Note that the reported performance per model in the tables is the average results via running 5 times with the best hyper-parameter settings.

\begin{equation}\label{eq:improv}
\begin{aligned}
&\text{Improvements} = \\
&\frac{\frac{\text{DCA-NARM}-\text{NARM}}{\text{NARM}}+\frac{\text{DCA-STAMP}-\text{STAMP}}{\text{STAMP}}+\frac{\text{DCA-GCEGNN}-\text{GCE-GNN}}{\text{GCE-GNN}}}{3}
\end{aligned}
\end{equation}

\begin{table*}[]
    \centering
    \caption{Model Performance on Diginetica. Best result is highlighted in boldface and the runner-up is underlined; `Improvements' indicates the average relative enhancements achieved by our DCA-SBRSs over the corresponding SBRSs as Equation~\ref{eq:improv}.}
    \vspace{-0.1in}
    \resizebox{\textwidth}{!}{  
    \begin{tabular}{c|cc|cc|cc|cc|cc|cc|cc}
        \toprule
         \multirow{2}{*}{Model$\backslash$Metric} & \multicolumn{2}{c}{NDCG} & \multicolumn{2}{c}{MRR} & \multicolumn{2}{c|}{HR} & \multicolumn{2}{c}{ILD} & \multicolumn{2}{c}{Entropy} & \multicolumn{2}{c|}{DS} & \multicolumn{2}{c}{F-score}\\\cline{2-15}
         & @10 & @20& @10 & @20& @10 & @20& @10 & @20& @10 & @20& @10 & @20& @10 & @20\\\midrule
         Item-KNN & 0.1313 & 0.1438 & 0.0999 & 0.1036 & 0.2343 & 0.2814 & 0.1653 & 0.2247 & 0.2852 & 0.4353 & 0.1562 & 0.1376 & 0.0375 & 0.0635\\ 
         BPR-MF & 0.0799 & 0.0954 & 0.0618 & 0.0661 & 0.1397 & 0.2012 & 0.5334 & 0.5799 & 0.9490 & 1.2148 & 0.2871 & 0.2159 & 0.0676 & 0.1061\\\hline
         NARM & 0.3191 & 0.3468 & 0.2578 & 0.2654 & \underline{0.5162} & \underline{0.6256} & 0.1811 & 0.2519 & 0.3047 & 0.5037 & 0.1575 & 0.1182 & 0.0921 & 0.1645 \\
         STAMP & 0.3143 & 0.3385 & 0.2558 & 0.2624 & 0.5018 & 0.5973 & 0.2704 & 0.3923 & 0.4781 & 0.8410 & 0.1977 & 0.1783 & 0.1381 & 0.2491  \\
         GCE-GNN & $\mathbf{0.3458}$ & $\mathbf{0.3723}$ & $\mathbf{0.2876}$ & $\mathbf{0.2950}$ & $\mathbf{0.5324}$ & $\mathbf{0.6373}$ & 0.1124 & 0.1623 & 0.1825 & 0.3096 & 0.1328 & 0.0892 & 0.0627 & 0.1145  \\\midrule
          MCPRN & 0.2321 & 0.2610 & 0.1858 & 0.1938 & 0.3829 & 0.4972 & 0.2671 & 0.3394 & 0.4651 & 0.7106 & 0.1935 & 0.1556 & 0.1100 & 0.1867  \\
          NARM+MMR & 0.2626 & 0.2896 & 0.2092 & 0.2167 & 0.4354 & 0.5420 & 0.3484 & 0.4574 & 0.6157 & 0.9691 & 0.2234 & 0.1909 & 0.1401 & 0.2443 \\
         IDSR($\lambda=0.8$) & 0.2681 & 0.2958 & 0.2140 & 0.2217 & 0.4438 & 0.5532 & 0.4105 & 0.4635 & 0.7464 & 1.0110 & 0.2593 & 0.2090 & 0.1814 &  0.2688\\\midrule
         DCA-NARM & 0.3226 & 0.3435 & 0.2641 & 0.2699 & 0.5099 & 0.5920 & \underline{0.4115} & \underline{0.6791} & \underline{0.7698} & \underline{1.6254} & \underline{0.2691} & \underline{0.3399} & \underline{0.2022} & \underline{0.4017}\\
         DCA-STAMP & 0.3067 & 0.3237 & 0.2529 & 0.2577 & 0.4779 & 0.5444 & $\mathbf{0.5750}$ & $\mathbf{0.8713}$ & $\mathbf{1.1009}$ & $\mathbf{2.1447}$ & $\mathbf{0.3489}$ & $\mathbf{0.4418}$ & $\mathbf{0.2693}$ & $\mathbf{0.4632}$ \\
         DCA-GCEGNN & \underline{0.3342} & \underline{0.3554} & \underline{0.2813} & \underline{0.2872} & 0.5032 & 0.5868 & 0.3090 & 0.5419 & 0.5844 & 1.2960 & 0.2304 & 0.2836 & 0.1426 & 0.3172\\\hline
         \emph{Improvements} & -1.56\% &  -3.29\% & -0.29\% & -0.91\% & -3.82\% & -7.38\% &  138\% &  175\% &   168\% &   232\% & 73.6\% &   184\% &   114\% &   136\% \\
         \bottomrule
    \end{tabular}}
    \label{tab:digi}
\end{table*}

\begin{table*}[htbp]
    \centering
    \caption{Model Performance on Retailrocket. Best result is highlighted in boldface and the runner-up is underlined; `Improvements' indicates the average relative enhancements achieved by our DCA-SBRSs over the corresponding SBRSs as Equation~\ref{eq:improv}.}
    \vspace{-0.1in}
    \resizebox{\textwidth}{!}{    
    \begin{tabular}{c|cc|cc|cc|cc|cc|cc|cc}
        \toprule
         \multirow{2}{*}{Model$\backslash$Metric} & \multicolumn{2}{c}{NDCG} & \multicolumn{2}{c}{MRR} & \multicolumn{2}{c|}{HR} & \multicolumn{2}{c}{ILD} & \multicolumn{2}{c}{Entropy} & \multicolumn{2}{c|}{DS} & \multicolumn{2}{c}{F-score}\\\cline{2-15}
         & @10 & @20& @10 & @20& @10 & @20& @10 & @20& @10 & @20& @10 & @20& @10 & @20\\\midrule
         Item-KNN & 0.1558 & 0.1634 & 0.1267 & 0.1289 & 0.2491 & 0.2777 & 0.6868 & 0.7954 & 1.2871 & 1.7206 & 0.3749 & 0.3822 & 0.1491 & 0.1979 \\
         BPR-MF & 0.1244 & 0.1369 & 0.1037 & 0.1072 & 0.1915 & 0.2407 & 0.8106 & 0.8599 & 1.5023 & 1.8863 & 0.4077 & 0.3183 & 0.1391 & 0.1899\\\hline
         NARM & 0.3625 & 0.3815 & 0.3138 & 0.3190 & 0.5181 & \underline{0.5928} & 0.4860 & 0.5885 & 0.8698 & 1.2658 & 0.2767 & 0.2369 & 0.2475 & 0.3507\\
         STAMP & 0.3516 & 0.3688 & 0.3068 & 0.3115 & 0.4945 & 0.5624 & 0.5313 & 0.6563 & 0.9769 & 1.4613 & 0.3046 & 0.2739 & 0.2530 & 0.3642\\
         GCE-GNN & $\mathbf{0.3917}$ & $\mathbf{0.4107}$ & $\mathbf{0.3426}$ & $\mathbf{0.3478}$ & $\mathbf{0.5481}$ & $\mathbf{0.6229}$ & 0.3701 & 0.4525 & 0.6312 & 0.9139 & 0.2207 & 0.1744 & 0.2143 & 0.3044\\\midrule 
         MCPRN & 0.2363 & 0.2501 & 0.2085 & 0.2123 & 0.3252 & 0.3799 & 0.7664 & 0.8432 & 1.4931 & 2.0162 & 0.4322 & 0.3852 & 0.2293 & 0.2930\\
         NARM+MMR & 0.3234 & 0.3413 & 0.2785 & 0.2834 & 0.4669 & 0.5375 & 0.6247 & 0.7436 & 1.1543 & 1.6684 & 0.3424 & 0.3073 & 0.2764 & 0.3863 \\
         IDSR($\lambda=0.5$) & 0.2863 & 0.3116 & 0.2526 & 0.2596 & 0.3998 & 0.4996 & $\mathbf{1.1929}$ & \underline{1.0939} & $\mathbf{2.4573}$ & \underline{2.7566} & $\mathbf{0.6794}$ & \underline{0.5506} & $\mathbf{0.4274}$ & \underline{0.5093} \\\midrule
         DCA-NARM & 0.3654 & 0.3804 & 0.3200 & 0.3241 & 0.5099 & 0.5688 & 0.7181 & 0.9328 & 1.3801 & 2.3049  & 0.4074  & 0.4618  & 0.3544 & 0.5053 \\
         DCA-STAMP & 0.3362 & 0.3471 & 0.2929 & 0.2960 & 0.4726 & 0.5155 & \underline{0.9061} & $\mathbf{1.1276}$ & \underline{1.8147} & $\mathbf{2.9613}$ & \underline{0.5230} & $\mathbf{0.6133}$ & \underline{0.3994} & $\mathbf{0.5257}$\\
         DCA-GCEGNN & \underline{0.3826} & \underline{0.3985} & \underline{0.3364} & \underline{0.3408} & \underline{0.5293} & 0.5921 & 0.5970 & 0.7813 & 1.1258 & 1.8713 & 0.3461 & 0.3754 & 0.3103 & 0.4533 \\\hline
         \emph{Improvements} & -1.97\% & -3.05\% & -1.45\% & -1.80\% & -3.15\% & -5.78\% &  59.9\% &  67.7\% &  74.3\% &  96.5\% & 58.6\% &  111\% &  48.6\% &  45.8\% \\
         \bottomrule
    \end{tabular}}
    \label{tab:updated_retail}
\end{table*}

\begin{table*}[htbp]
\caption{Model Performance on Tmall. Best result is highlighted in boldface and the runner-up is underlined; `Improvements' indicates the average relative enhancements achieved by our DCA-SBRSs over the corresponding SBRSs as Equation~\ref{eq:improv}.}\label{tab:tmall}
\centering
    \vspace{-0.1in}
    \resizebox{\textwidth}{!}{  
    \begin{tabular}{c|cc|cc|cc|cc|cc|cc|cc}
        \toprule
         \multirow{2}{*}{Model$\backslash$Metric} & \multicolumn{2}{c}{NDCG} & \multicolumn{2}{c}{MRR} & \multicolumn{2}{c|}{HR} & \multicolumn{2}{c}{ILD} & \multicolumn{2}{c}{Entropy} & \multicolumn{2}{c|}{DS} & \multicolumn{2}{c}{F-score}\\\cline{2-15}
         & @10 & @20& @10 & @20& @10 & @20& @10 & @20& @10 & @20& @10 & @20& @10 & @20\\\midrule
         Item-KNN & $\mathbf{0.0321}$ & $\mathbf{0.0349}$ & $\mathbf{0.0251}$ & $\mathbf{0.0259}$ & {0.0551} & {0.0655} & 0.8888 & 0.9593 & 1.6790 & 2.0452 & 0.4546 & 0.4219 & {0.0442} &  0.0573\\ 
         BPR-MF & 0.0096 & 0.0119 & 0.0069 & 0.0075 & 0.0186 & 0.0279 & 0.9963 & 1.0350 & 1.8716 & 2.3219 & 0.4852 & 0.3805 & 0.0168 &  0.0259\\\hline
         NARM & 0.0244 & 0.0306 & {0.0174} & {0.0191} & 0.0476 & 0.0720 & 0.9453 & 1.0085 & 1.7760 & 2.2625 & 0.4689 & 0.3778 & 0.0386 &  0.0642\\ 
         STAMP & 0.0171 & 0.0215 & 0.0121 & 0.0133 & 0.0336 & 0.0511 & 1.0449 & 1.0959 & 2.0375 & 2.5806 & 0.5428 & 0.4494 & 0.0292 &  0.0481\\
         GCE-GNN & \underline{0.0282} & \underline{0.0355} & \underline{0.0187} & \underline{0.0207} & $\mathbf{0.0594}$ & $\mathbf{0.0886}$ & 0.8571 & 0.9326 & 1.5691 & 2.0340 & 0.4161 & 0.3345 & \underline{0.0443} & \underline{0.0744} \\\midrule
         MCPRN & 0.0110 & 0.0142 & 0.0075 & 0.0084 & 0.0225 & 0.0354 & 1.0661 & 1.1042 & 2.1139 & 2.6437 & 0.5686 & 0.4679 & 0.0193 & 0.0326 \\
         NARM+MMR & 0.0198 & 0.0249 & 0.0141 & 0.0154 & 0.0386 & 0.0592 & 1.0116 & 1.0634 & 1.9386 & 2.4437 & 0.5124 & 0.4155 & 0.0331 & 0.0548 \\
         IDSR($\lambda=0.8$) &0.0083 & 0.0114 & 0.0054 & 0.0063 & 0.0179 & 0.0303 & $\mathbf{1.3175}$ & 1.2969 & \underline{2.8725} & 3.4530 & \underline{0.8108} & 0.6773 & 0.0192 & 0.0327 \\\midrule
         DCA-NARM & 0.0145 & 0.0171 & 0.0106 & 0.0113 & 0.0272 & 0.0374 & \underline{1.3096} & $\mathbf{1.3466}$ & $\mathbf{2.9308}$ & $\mathbf{3.8986}$ & $\mathbf{0.8548}$ & $\mathbf{0.8566}$ & 0.0274 & 0.0402 \\
         DCA-STAMP & 0.0164 & 0.0192 & 0.0122 & 0.0129 & 0.0304 & 0.0414 & 1.2720 & \underline{1.3274} & 2.7719 & \underline{3.7395} & 0.7919 & \underline{0.7962} & 0.0311 & 0.0453\\
         DCA-GCEGNN & {0.0259} & {0.0329} & 0.0165 & 0.0185 & \underline{0.0566} & \underline{0.0843} & 0.9647 & 1.0464 & 1.8368 & 2.4230 & 0.4894 & 0.4205 & $\mathbf{0.0467}$ & $\mathbf{0.0777}$\\\hline
         \emph{Improvements.} & -17.6\% & -20.7\% & -16.7\% & -18.16\% & -19.03\% & -24.0\% & 24.3\% &  22.3\% &  39.4\% &  45.4\% & 48.6\% &  76.5\% & -5.70\% & -12.9\% \\
         \bottomrule
    \end{tabular}}
\end{table*}

\subsubsection{Performance on Recommendation Accuracy}
The accuracy of all approaches is measured via NDCG@$N$, MRR@$N$, and HR@$N$ ($N=\{10, 20\}$) in Tables~\ref{tab:digi}-\ref{tab:tmall}, where several observations are obtained as follows.
\textbf{1)} For traditional methods, Item-KNN outperforms BPR-MF across all three datasets.
Both are generally defeated by the deep neural approaches, except for Item-KNN on Tmall.
\textbf{2)} 
Compared with our proposed framework, the existing accuracy-focused SBRSs come in first with the help of the neural network to learn more precise item embeddings and attention mechanism to denoise. Among them, GCE-GNN outperforms other methods on all three datasets, which demonstrates the expressive power of local current session graph and global session graph.
\textbf{3)} The accuracy of the aforementioned SBRSs is slightly decreased under our DCA framework, with few exceptions, such as DCA-NARM vs. NARM on Diginetica and Retailrocket. While, the perturbation (e.g., with $1.6\%$ and $2\%$ drops on average w.r.t. NDCG@$10$ on Diginetica and Retailrocket respectively) can be tolerated given our significant enhancements in diversity and comprehensive metrics, which will be elaborated in what follows.
\textbf{4)} Deep diversified SBRSs generally perform better than traditional methods whereas worse than the accuracy-oriented deep methods due to their special design for gaining higher diversity. 
{In contrast to NARM, the accuracy of NARM+MMR drops significantly across three datasets.} It's worth noting that our DCA-SBRSs show a superior advantage over deep diversified methods, for instance, the performance of our DCA-SBRS is one time better than IDSR w.r.t. HR@$10$ on Tmall.

\subsubsection{Performance on Recommendation Diversity}
The diversity of all comparisons is measured via ILD@$N$, Entropy@$N$, and DS@$N$ ($N=\{10, 20\}$) in Tables~\ref{tab:digi}-\ref{tab:tmall}.
Three major findings can be noted. \textbf{1)} Existing SBRSs benefit significantly from our proposed DCA framework. For instance, averagely, across the three datasets, the relative improvements regarding diversity on ILD@$10$ achieved by our DCA-SBRSs over the corresponding SBRSs (e.g., DCA-NARM vs. NARM) can reach $138\%$, $59.9\%$, and $24.3\%$, respectively. Besides, some of our DCA-SBRSs (e.g., DCA-STAMP) outperform all other methods (including the deep diversified models) on Diginetica and Tmall.
\textbf{2)} Towards diversified models, the performance of IDSR exceeds that of MCPRN on all three datasets. Meanwhile, 
{all} of them beat existing accuracy-oriented SBRSs (except MCPRN vs. STAMP on Diginetica), indicating the efficacy of these diversified methods in gaining better diversity.
\textbf{3)} Existing accuracy-oriented SBRSs perform worst due to ignoring the demands on diversity. Among them, STAMP performs best across all three datasets. Moreover, 
traditional methods (led by BPR-MF), though being surpassed by these accuracy-oriented SBRSs with regard to recommendation accuracy,
perform slightly better when it comes to diversity. 

\subsubsection{Comprehensive Performance}\label{subsub:comprehensive}
To comprehensively assess the performance from both accuracy and diversity perspectives, we further compare them in terms of F-score@$N$ ($N=\{10, 20\}$) in Tables~\ref{tab:digi}-\ref{tab:tmall}, and several interesting findings can be gained. \textbf{1)} Our proposed DCA-SBRSs perform the best among all baselines. Specifically, a quite encouraging phenomenon is observed that some of our DCA-SBRSs 
{show effectiveness by} defeating diversified models in terms of both accuracy and diversity (e.g., DCA-STAMP on Diginetica, DCA-NARM on Tmall 
{and DCA-NARM vs. NARM+MMR on Diginetica and Retailrocket}). Additionally, our framework also outperforms accuracy-oriented SBRSs with significant gains on diversity while only minor drops on accuracy.
\textbf{2)} Towards deep diversified models, IDSR achieves both better accuracy and diversity than MCPRN and NARM+MMR on Diginetica and Retailrocket, demonstrating the superiority of IDSR against MCPRN. 
\textbf{3)} Typically, traditional methods perform worse than accuracy-oriented SBRSs. Comparing accuracy-oriented SBRSs and diversified SBRSs, the former performs 
better on Tmall, while worse on Diginetica. 
This is mainly caused by the calculation of the F-score (harmonic mean of HR and ILD). 
Due to the different features (e.g., distribution) of various datasets, the results achieved on different datasets regarding HR and ILD may vary a lot. For instance, the ILD values are generally higher than HR values on Diginetica, while the opposite case is held on Tmall.
Therefore, the model achieving the best result w.r.t. the weaker metric (e.g., HR on Diginetica) will gain advantages regarding the comprehensive performance, i.e., F-score.

Interestingly, we notice that all methods perform worse regarding the recommendation accuracy whilst better w.r.t. diversity on Tmall compared with the other two datasets. This might be caused by the unique data distribution of Tmall, i.e., lower RR and higher DS in Table~\ref{tab:updated_datasetStatistics}. Nevertheless, our proposed DCA still exceeds other diversified SBRSs, showing the stability of our DCA.


\begin{table}[t]
\centering
\caption{Computational Time Comparison.}\label{tab:time_complexity}
\addtolength{\tabcolsep}{-1pt}
{
\begin{tabular}{l|r|r|r|r|r|r}\toprule
\multirow{2}{*}{Model\textbackslash{}Time} & \multicolumn{3}{|c|}{Training(/epoch)} & \multicolumn{3}{c}{Inference}      \\\cline{2-7}
                                           & Digi*  & Retail*  & Tmall  & Digi*  & Retail* & Tmall \\\midrule
NARM                                       & 49s       & 68s        &  156s      & 560s     & 137s      &   2678s    \\
NARM+MMR                                   & 49s       & 68s        &  156s      & 5173s & 2082s     &  6075s     \\
MCPRN       & 244s           &  3003s             &  1138s      &  2325s           &   1689s       &   3167s    \\
IDSR                                       & 1486s    &   1646s            &   4604s     & 62s      &  29s            &  1928s    \\
DCA-NARM                                   & 127s      & 159s       &  418s      & 8s        & 4s             &  134s     \\\bottomrule
\multicolumn{7}{l}{\footnotesize Note: Diginetica and Retailrocket are shortened as Digi* and Retail*.} \\
\end{tabular}
}
\vspace{-0.1in}
\end{table}

\subsubsection{Performance on Time Complexity}
Following the discussion in Section~\ref{subsec:simple}, we empirically verify the efficiency of our lightweight DCA. As such, we record the training and inference time
for representative methods, including NARM, NARM+MMR, DCA-NARM, MCPRN, IDSR and our DCA-NARM, 
across three datasets shown in Table~\ref{tab:time_complexity}. Two major findings are noted. \textbf{1)}  
MMR is a re-ranking (two-stage) method by a greedy search for diversity-promoting based on the trained NARM from the first step (training stage). NARM+MMR hence has a substantially longer inference time than NARM. By contrast, our DCA+NARM accomplishes an end-to-end learning and avoids greedy search in the inference stage, thus being faster than NARM+MMR. \textbf{2)} Unlike other diversified SBRSs (i.e., IDSR and MCPRN) relying on specifically calibrated diversity-aware components, our DCA framework performs effectively on both training and inference stages due to limited additional parameters.



\subsubsection{Adaptation on F-score}
We now discuss the drawbacks of the current comprehensive metric (F-score~\cite{hu2017diversifying}), and provide remedies accordingly. First, due to different scales of HR and ILD, the weaker metric may easily dominate the final comprehensive performance, particularly on Tmall in Table~\ref{tab:tmall}. 
Therefore, it is necessary to map the two metrics into the same range before calculating F-score. Alternatively, we may replace ILD with DS (Diversity Score~\cite{liang2021enhancing}) in F-score since HR and DS are in the same range of $[0, 1]$. 
Second, \textit{\textbf{a clear decline on accuracy is generally not acceptable in real-world recommendation scenarios}}. According to Tables~\ref{tab:digi}-\ref{tab:tmall}, diversified models have apparent drops on accuracy due to the significant improvements on diversity. However, their comprehensive performance (i.e., F-score) is not the worst, even the best on Retailrocket (Table~\ref{tab:updated_retail}). That is to say, the current comprehensive performance does not match what is actually anticipated by the real-world applications. As such, we propose a generalized comprehensive metric  F$_{\beta}(\text{ACCuracy}, \text{DIVersity})$ to solve the aforementioned issue, as below:
\begin{equation}
\text{F}_{\beta}\text{(ACC, DIV)}=\frac{ (1+\beta^2) \text{ACC}\times \text{DIV}}{\beta^2\text{ACC}+\text{DIV}}\label{eq:F-score},
\end{equation}
where $\beta>0$. Accordingly, the F-score~\cite{hu2017diversifying} can be regarded as a special case, i.e., $\text{F}_{1}\text{(HR, ILD)}$. For a consistent range of ACC and DIV, we recommend $\text{F}_{\beta}\text{(HR, DS)}$. Additionally,
if accuracy is prioritized over diversity, 
we suggest $\beta<1$, e.g., $\text{F}_{0.5}\text{(HR, DS)}$, to put more emphasis on accuracy since it is less meaningful to gain diversity without taking accuracy into account in real-world applications. 
Note that with the proposed $\text{F}_{\beta}\text{(ACC, DIV)}$, our proposed DCA-SBRSs rank first thanks to the satisfying performance on accuracy and superior performance on diversity, as shown in Table~\ref{tab:updated_retail_f_beta}. Specifically, on Retailrocket, the ranking of our DCA-SBRS improves w.r.t. $N=10$, while diversified models (e.g., IDSR) experience a decline in ranking by changing $\beta$ from $1$ to $0.5$ due to its inferior accuracy performance.

\begin{table}[t]
    \centering
    \caption{F-score vs. {Adapted F-score} on Retailrocket. Best result is highlighted in boldface and the runner-up is underlined; `Improvements' indicates the average relative enhancements achieved by our DCA-SBRSs over the corresponding SBRSs as Equation~\ref{eq:improv}.}
    \begin{adjustbox}{max width=\linewidth}  
    \begin{tabular}{c|cc|cc|cc}
        \toprule
         \multirow{2}{*}{Model$\backslash$Metric} & \multicolumn{2}{c|}{F-score} & \multicolumn{2}{c|}{F$_{0.5}$(HR,ILD)} & \multicolumn{2}{c}{F$_{0.5}$(HR,DS)}\\\cline{2-7}
         & @10 & @20& @10 & @20& @10 & @20\\\midrule
         Item-KNN & 0.1491 & 0.1979 & 0.1618 & 0.2093 & 0.1542 & 0.1714\\
         BPR-MF & 0.1391 & 0.1899 & 0.1452 & 0.1980 & 0.1287 & 0.1418\\\hline
         NARM &0.2475 & 0.3507 & 0.2782 & 0.3960 & 0.2888 & 0.2938\\
         STAMP & 0.2530 & 0.3642 & 0.2790 & 0.4008 & 0.2859 & 0.3024\\
         GCE-GNN & 0.2143 & 0.3044 & 0.2439 & 0.3544 & 0.2820 & 0.2681\\\midrule 
         MCPRN & 0.2293 & 0.2930 & 0.2393 & 0.3059 & 0.2193 & 0.2349\\
         NARM+MMR & 0.2764 & 0.3863 & 0.3013 & 0.4174 & 0.2871 & 0.3110\\
         IDSR($\lambda=0.5$) & $\mathbf{0.4274}$ & \underline{0.5093} & \underline{0.4095} & 0.5011 & \underline{0.3573} & \underline{0.4142} \\\midrule
         DCA-NARM & 0.3544 & 0.5053 & 0.3787 & $\mathbf{0.5180}$ & 0.3447 & 0.4138\\
         DCA-STAMP & \underline{0.3994} & $\mathbf{0.5257}$ & $\mathbf{0.4122}$ & \underline{0.5128} & $\mathbf{0.3606}$ & $\mathbf{0.4319}$\\
         DCA-GCEGNN & 0.3103 & 0.4533 & 0.3391 & 0.4838 & 0.3269 & 0.3761\\\hline
         \emph{Improvements.} & 48.6\% &  45.8\%  & 41.0\%  & 31.8\%  & 20.5\%  & 41.3\%  \\
         \bottomrule
    \end{tabular}
    \end{adjustbox}
    \label{tab:updated_retail_f_beta}
\end{table}

\subsection{The Impact of Essential Modules}

\subsubsection{Impact of Model-agnostic Diversified Loss (abbr. {MDL})}\label{subsec:impact_of_DL}

\begin{figure}[t]
    \centering
    \includegraphics[width=.5\textwidth]{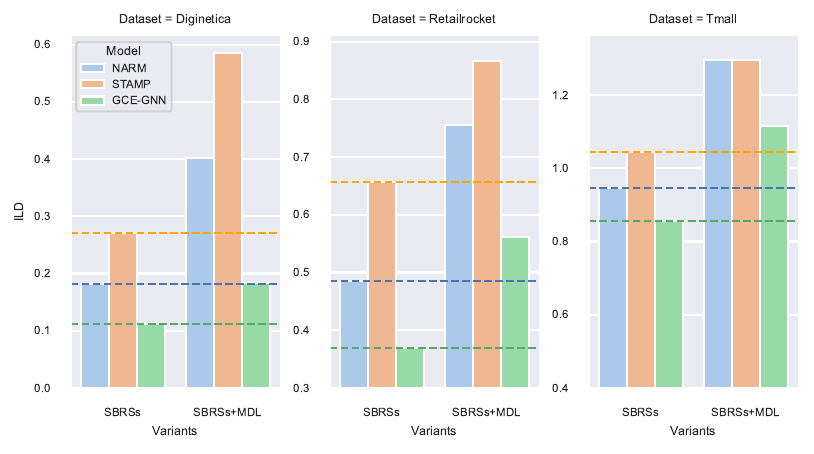}    
    \vspace{-0.15in}
    \caption{The Impact of 
    {MDL} in Diversity w.r.t. ILD@$10$.}
    \label{fig:ablation_dl_ild}
    \vspace{-0.1in}
\end{figure}

Our proposed MDL in Equation~\ref{equ:l_div} aims to improve the diversity of accuracy-oriented SBRSs as an end-to-end plugin 
{by punishing monotonous RL with low diversity}. In Figure~\ref{fig:ablation_dl_ild}, we compare the accuracy-oriented SBRSs (labeled as `SBRSs') and
the corresponding variants with our {MDL} supplemented solely (labeled as `SBRSs+{MDL}') w.r.t. 
ILD@$10$. Accordingly, by adding {our MDL}, the diversity of all baseline SBRSs significantly improves across the three datasets. Specifically, on Diginetica, Retailrocket, and Tmall, the average relative improvements are $100\%$, $56.7\%$, and $30.3\%$, respectively. Besides, among the three selected baselines (NARM, STAMP, and GCE-GNN), {MDL} improves NARM most (i.e., $71.46\%$).

\begin{figure}[t]
    \centering
    \includegraphics[width=.5\textwidth]{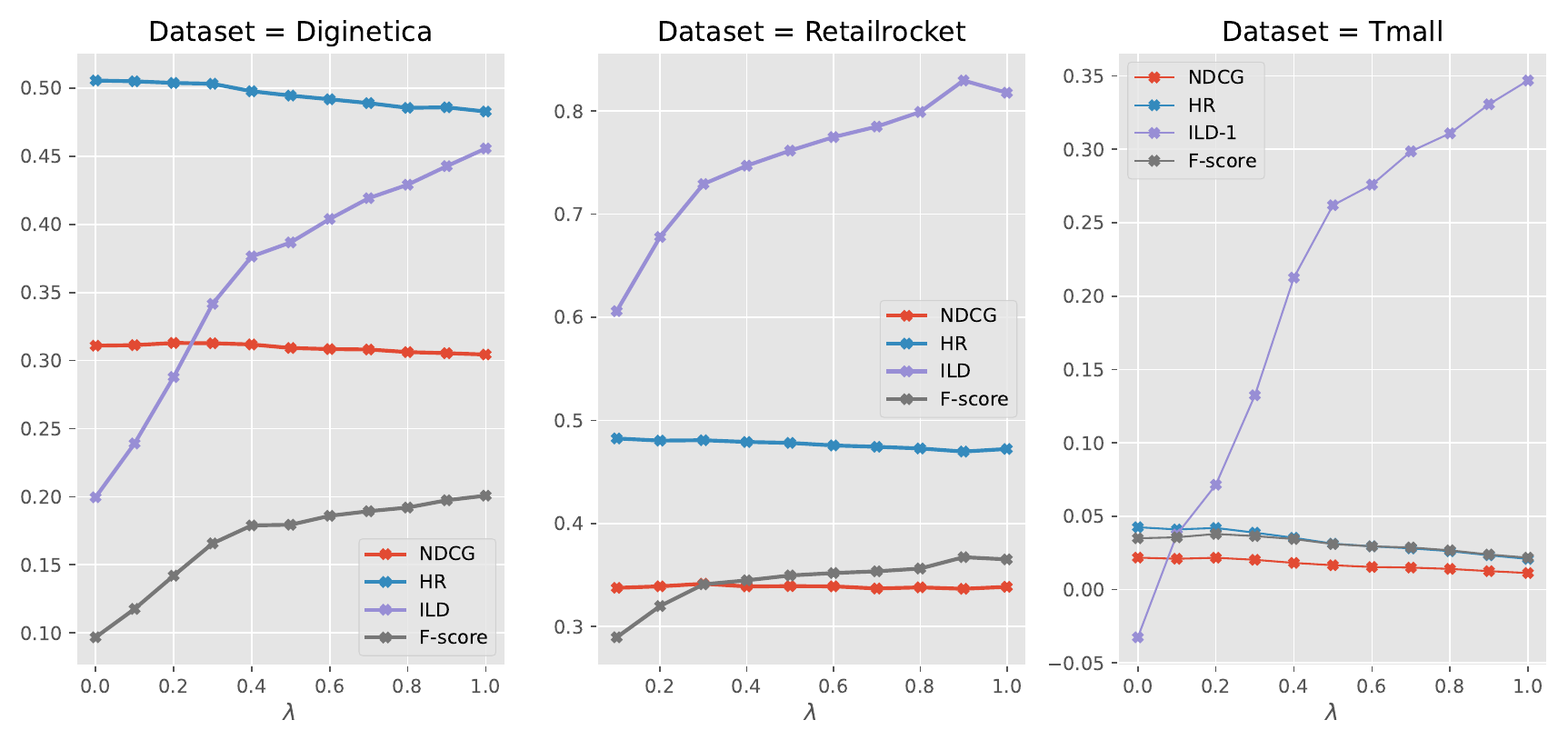}
    \vspace{-0.15in}
    \caption{The Impact of 
    {MDL} for NARM+{MDL} with $N=10$.}
    \label{fig:dl_fined}
    \vspace{-0.1in}
\end{figure}

It's worth noting that, for simplicity, we set $\lambda=1$ in Equation~\ref{equ:final-loss}. To analyze the effect of MDL in a fine-grained manner, we select NARM as our basic predictor and vary the value of $\lambda$ from 0 to 1 stepped by 0.1. 
Figure~\ref{fig:dl_fined} depicts the variation w.r.t. accuracy (i.e., NDCG and HR), diversity (i.e., ILD), and comprehensive performance (i.e., F-score) with varied $\lambda$ on the three datasets\footnote{For ease of presentation, we display the values of `ILD minus one' (i.e., ILD-1) on Tmall to ensure all metrics in a proper scale without changing the overall trend.}.  
As noted, the \textit{\textbf{accuracy}} slightly decreases with the increasing of $\lambda$ on all three datasets; whilst a significant enhancements on \textit{\textbf{diversity}} is noted on all datasets, showcasing the remarkable effectiveness of our MDL. 
Towards \textit{\textbf{comprehensive performance}}, F-score climbs up when $\lambda$ varies from 0 to 1 on Diginetica and Retailrocket; whereas it has a slight decline on Tmall. The possible explanation can be found in Section~\ref{subsub:comprehensive}. 
As a whole, the recommendation accuracy drops and diversity increases by boosting the value of $\lambda$ gradually. This indicates the necessity of fune-tuning $\lambda$ to achieve more satisfying performance.

\subsubsection{Impact of {Non-invasive} Category-aware Attention (abbr. {NCA})}

\begin{figure}[t]
    \centering
    \includegraphics[width=.5\textwidth]{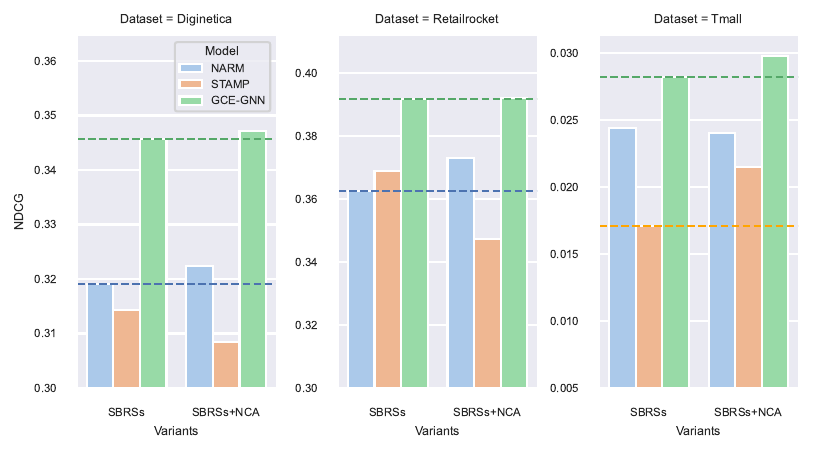}
    \vspace{-4mm}
    \caption{The Impact of NCA in Accuracy w.r.t. NDCG@$10$.}
    \label{fig:ablation_ca_ndcg}
    \vspace{-0.1in}
\end{figure}

As indicated in Section~\ref{subsec:impact_of_DL}, the recommendation accuracy of baseline SBRSs may slightly drops when integrating our designed {MDL}. To ease this issue, we propose category-aware attention (i.e., {NCA}) by importing category information into the pervasive attention mechanisms in SBRSs, with the goal of assisting item prediction. This differs from simply concatenating category information as the input of SBRSs. 
For verification, we compare accuracy-oriented SBRSs (labeled as `SBRSs') and the corresponding variants by simply substituting the attention mechanism with our category-aware attention (labeled as `SBRSs+{NCA}') on accuracy (i.e., NDCG@10), as depicted in
Figure~\ref{fig:ablation_ca_ndcg}.
In general, replacing the attention mechanism with our {NCA} facilitates the accuracy of SBRSs. Specifically, {NCA} helps NARM and GCE-GNN enhance their accuracy on all datasets. A similar trend is held by STAMP on Tmall; however, on the other two datasets, the accuracy of STAMP+{NCA} has not improved. That is perhaps due to the straightforward design of STAMP, which employs item embeddings directly rather than hidden states from RNNs or GNNs (e.g., NARM and GCE-GNN). As a result, STAMP+{NCA} simply sums item embeddings and the relevant category embeddings before computing attention scores, which may introduce more noise to interfere with the final item prediction. 

There's no denying that our DCA framework aids existing accuracy-oriented SBRSs in achieving extraordinary diversity and comprehensive performance gains while maintaining accuracy simultaneously, even 
without a thorough accuracy improvements 
for all SBRSs+{NCA} on all datasets as shown in Figure~\ref{fig:ablation_ca_ndcg} (this may be caused by different features of datasets or designs of baseline predictors). Alternatively stated, the efficacy of our proposed framework does not rely on {NCA} only. 

\subsection{Discussion on our proposed DCA framework}
Our proposed Diversified Category-aware Attentive (DCA) framework comprises two key components: a model-agnostic diversity-oriented loss function and a non-invasive category-aware attention mechanism. To evaluate the efficacy of the DCA framework, we selected three deep neural methods with attention mechanisms as their backbone, as detailed in Section \ref{subsec:baselines} and Section \ref{sec:results}. Notably, these methods all rank among the top five SBRSs in terms of accuracy \cite{under2023}. In the session-based evaluation survey \cite{under2023}, it is evident that all of the top-performing SBRSs in accuracy leverage attention mechanisms.

\begin{figure}[htb]
	\centering
	\begin{tikzpicture}
	\begin{groupplot}[group style={
		group name=myplot,
		group size= 2 by 3,  horizontal sep=1.0cm}, height=4.5cm,width=5cm,
	ylabel style={yshift=-0.6cm},
	legend style = {font=\tiny, column sep=-1cm},
	every tick label/.append style= {font=\tiny}
	]
	\nextgroupplot[ybar=0.10,
	bar width=0.45em,
	ylabel={HR@10},
	scaled ticks=false,
	yticklabel style={/pgf/number format/.cd,fixed,precision=3},
  	ymin=0, 
	enlarge x limits=0.25,
	symbolic x coords={Diginetica,Retailrocket,Tmall},
	ylabel style =  {font=\tiny},
	xtick=data,
        legend to name=CombinedLegendBar,
	]
	\addplot[color=teal, fill=teal, fill opacity=0.65] coordinates {
		(Diginetica,0.2141) (Retailrocket, 0.3467)  (Tmall, 0.0068)};\label{plots:plot1}
  	\addplot[color=pink, fill=pink, opacity=0.65] coordinates {
		(Diginetica,0.5162) (Retailrocket, 0.5181) (Tmall,0.0476)};\label{plots:plot2}
	\addplot[color=purple, fill=purple, opacity=0.65] coordinates {
		(Diginetica,0.5099) (Retailrocket, 0.5099) (Tmall,0.0272)};\label{plots:plot3}

	\nextgroupplot[ybar=0.1,
	bar width=0.45em,
	ylabel={HR@20},
	scaled ticks=false,
	yticklabel style={/pgf/number format/.cd,fixed,precision=3},
  	ymin=0, 
	enlarge x limits=0.25,
	symbolic x coords={Diginetica,Retailrocket,Tmall},
	xtick=data,
	ylabel style = {font=\tiny},
        legend to name=CombinedLegendBar,]
	\addplot[color=teal, fill=teal, fill opacity=0.65] coordinates {
		(Diginetica,0.2837) (Retailrocket, 0.3984)   (Tmall, 0.0097)};
    	\addplot[color=pink, fill=pink, opacity=0.65] coordinates {
		(Diginetica,0.6256) (Retailrocket, 0.5928) (Tmall,0.0720)};
	\addplot[color=purple, fill=purple, opacity=0.65] coordinates {
		(Diginetica,0.5920) (Retailrocket, 0.5688)  (Tmall, 0.0374)};

	\nextgroupplot[ybar=0.1,
	bar width=0.45em,
	ylabel={ILD@10},
	scaled ticks=false,
	yticklabel style={/pgf/number format/.cd,fixed,precision=3},
 	ymin=0, 
	enlarge x limits=0.25,
	symbolic x coords={Diginetica,Retailrocket,Tmall},
	xtick=data,
	ylabel style =  {font=\tiny},
        legend to name=CombinedLegendBar,]
	\addplot[color=teal, fill=teal, fill opacity=0.65] coordinates {
		(Diginetica,0.8351) (Retailrocket, 1.0513) (Tmall, 1.3415)};
        \addplot[color=pink, fill=pink, opacity=0.65] coordinates {
        (Diginetica,0.1811) (Retailrocket, 0.4860) (Tmall,0.9453)};
	\addplot[color=purple, fill=purple, opacity=0.65] coordinates {
		(Diginetica,0.4115) (Retailrocket, 0.7181)  (Tmall, 1.3096)};

  	\nextgroupplot[ybar=0.1,
	bar width=0.45em,
	ylabel={ILD@20},
	scaled ticks=false,
	yticklabel style={/pgf/number format/.cd,fixed,precision=3},
  	ymin=0, 
	enlarge x limits=0.25,
	symbolic x coords={Diginetica,Retailrocket,Tmall},
	xtick=data,
	ylabel style = {font=\tiny},
        legend to name=CombinedLegendBar,]
	\addplot[color=teal, fill=teal, fill opacity=0.65] coordinates {
    (Diginetica, 0.9126) (Retailrocket, 1.1498) (Tmall, 1.3517)};
      	\addplot[color=pink, fill=pink, opacity=0.65] coordinates {
		(Diginetica,0.2519) (Retailrocket, 0.5885) (Tmall,1.0085)};
    \addplot[color=purple, fill=purple, opacity=0.65] coordinates {
    (Diginetica, 0.6791) (Retailrocket,  0.9328) (Tmall, 1.3466)};

	\nextgroupplot[ybar=0.1,
	bar width=0.45em,
	ylabel={F-score@10},
	scaled ticks=false,
	yticklabel style={/pgf/number format/.cd,fixed,precision=3},
  	ymin=0, 
	enlarge x limits=0.25,
	symbolic x coords={Diginetica,Retailrocket,Tmall},
	xtick=data,
	ylabel style =  {font=\tiny},
        legend to name=CombinedLegendBar,]
    \addplot[color=teal, fill=teal, fill opacity=0.65] coordinates {
    (Diginetica, 0.0970) (Retailrocket, 0.3075) (Tmall, 0.0075)};
      	\addplot[color=pink, fill=pink, opacity=0.65] coordinates {
		(Diginetica,0.0921) (Retailrocket, 0.2475) (Tmall,0.0386)};
    \addplot[color=purple, fill=purple, opacity=0.65] coordinates {
    (Diginetica, 0.2022) (Retailrocket, 0.3544) (Tmall,0.0274)};

	\nextgroupplot[ybar=0.1,
	bar width=0.45em,
	ylabel={F-score@20},
	scaled ticks=false,
	yticklabel style={/pgf/number format/.cd,fixed,precision=3},
  	ymin=0, 
	enlarge x limits=0.25,
	symbolic x coords={Diginetica,Retailrocket,Tmall},
	xtick=data,
	ylabel style = {font=\tiny},
        legend to name=CombinedLegendBar,]
	\addplot[color=teal, fill=teal, fill opacity=0.65] coordinates {
    (Diginetica, 0.1635) (Retailrocket, 0.3984) (Tmall, 0.0109)};
      	\addplot[color=pink, fill=pink, opacity=0.65] coordinates {
		(Diginetica,0.1645) (Retailrocket, 0.3507) (Tmall,0.0642)};
    \addplot[color=purple, fill=purple, opacity=0.65] coordinates {
    (Diginetica, 0.4017) (Retailrocket,  0.5053) (Tmall, 0.0402)};
	\end{groupplot}
    \path (myplot c2r1.north west)--
      coordinate(legendpos)
      (myplot c1r1.north east);
\node[
    matrix of nodes,
    draw,
    inner sep=0.2em,
    font=\tiny,
]at([yshift=2ex]legendpos)
  {
    \ref{plots:plot1} & GRU4Rec & [1em]
    \ref{plots:plot2} & GRU4Rec+Attention & [1em]
    \ref{plots:plot3} & DCA-GRU4Rec \\};
\end{tikzpicture}
\caption{Performance comparison between GRU4Rec and two variants in terms of HR, ILD, and F-score $N\in\{10,20\}$.}\label{fig:gru_dca}
\end{figure}
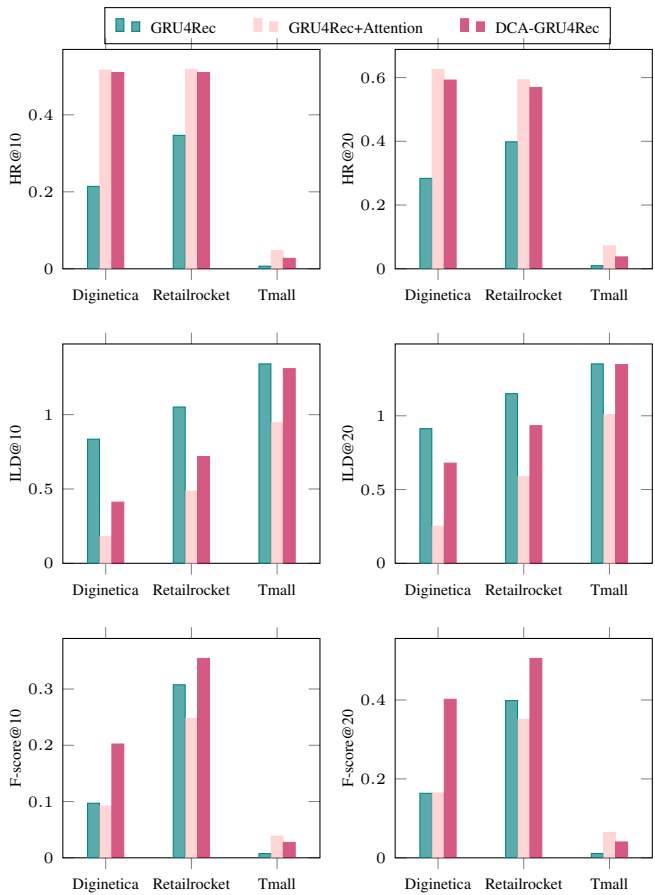

However, our DCA framework isn't limited solely to attention-based models. Despite the original SBRS not making use of an attention mechanism, we demonstrate the seamless integration of this component for enhanced session representation. Specifically, we adopt GRU4Rec\cite{hidasi2015session}, an RNN-based SBRS without an attention mechanism, as our backbone model to showcase the effectiveness of our DCA framework in this context.
As illustrated in Figure~\ref{fig:gru_dca}, we compare GRU4Rec with two variants: GRU4Rec with an attention mechanism and DCA-GRU4Rec, considering accuracy, diversity, and comprehensive performance. In summary, GRU4Rec with an attention mechanism outperforms the baseline GRU4Rec in terms of accuracy but lags in terms of diversity. Our DCA-GRU4Rec, on the other hand, achieves similar accuracy to GRU4Rec with an attention mechanism while significantly enhancing diversity and delivering a satisfactory overall performance. This substantiates the effectiveness of our DCA framework when applied to backbone models without attention mechanisms.

In conclusion, our DCA framework is highly versatile and can be seamlessly integrated into common SBRSs, whether they incorporate attention mechanisms or not, consistently showcasing its effectiveness in enhancing recommendation system performance.


\section{Conclusions}
In this paper, we have proposed a simple yet effective diversified category-aware attentive SBRSs (DCA-SBRSs) to improve diversity over SOTA accuracy-oriented SBRSs in the meantime striving to maintain the recommendation accuracy. To fulfill the goals, our DCA-SBRS consists of two novel components: (1) a model-agnostic diversity-oriented loss  function for diversity purpose; and (2) a non-invasive category-aware attention mechanism, which exploits category information for SBRS in a non-invasive way to keep accuracy of original SBRSs.
Our generic framework can serve as a plugin and be easily instantiated with representative accuracy-oriented SBRSs. 
Extensive experiments on three datasets show that: (1) DCA-SBRSs significantly outperform the corresponding baselines in terms of diversity and comprehensive performance while maintaining satisfying accuracy performance; (2) both the two components are effective in terms of the respective goal. Moreover, we have  discussed the limitations of existing comprehensive performance metrics considering both accuracy and diversity, and offered more reasonable strategy to evaluate diversified recommenders.



\bibliographystyle{unsrt}  
\bibliography{sample-base}

\end{document}